\documentclass[aps,twocolumn]{revtex4}
\usepackage[utf8]{inputenc}
 \usepackage{amsmath}
\usepackage{amssymb}
\usepackage{latexsym}
\usepackage{epsfig}
\usepackage{color}

\usepackage{amsmath,amssymb}
\usepackage{graphicx} % Allows including images
\usepackage{gastex}
\usepackage{epstopdf}
\usepackage[colorlinks=true,citecolor=blue,linkcolor=blue,urlcolor=black]{hyperref}
\usepackage{color}
\graphicspath{{Images/}}

\renewcommand{\&}{\textup{\symbol{`\&}}}
\begin{document}
\title{Holographic dark energy in modified Barrow cosmology}

\author{Ahmad Sheykhi}
\email{asheykhi@shirazu.ac.ir} \affiliation{Department of Physics,
College of Sciences, Shiraz University, Shiraz 71454, Iran}
\affiliation{Biruni Observatory, College of Sciences, Shiraz
University, Shiraz 71454, Iran}

\author{Maral Sahebi Hamedan}
\affiliation{Department of Physics, College of Sciences, Shiraz
University, Shiraz 71454, Iran} \affiliation{Biruni Observatory,
College of Sciences, Shiraz University, Shiraz 71454, Iran}

\begin{abstract}
Thermodynamics-Gravity conjecture implies that there is a deep
connection between the gravitational field equations and the first
law of thermodynamics. Therefore, any modification to the entropy
expression, directly modifies the field equations. By considering
the modified Barrow entropy associated with the apparent horizon,
the Friedmann equations get modified as well. In this paper, we
reconsider the holographic dark energy (HDE) model when the
entropy is in the form of Barrow entropy. This modification to the
entropy not only changes the energy density of HDE, but also
modifies the Friedmann equations. Therefore, one should take into
account the modified HDE in the context of modified Friedmann
equations. We study the Hubble horizon and the future event
horizon as IR cutoffs, and investigate the cosmological
consequences of this model. We also extend our study to the case
where dark matter (DM) and dark energy (DE) interact with each
other. We observe that Barrow exponent $\delta$ significantly
affects the cosmological behavior of HDE, and in particular the
equation of state (EoS) parameter can cross the phantom line
$(w_{de}<-1)$. Also adding $\delta$ remarkably affects the
deceleration parameter, and shifts the time of universe phase
transition.
\end{abstract}

\maketitle

\section{Introduction}
A huge number of cosmological observations support that our
Universe is currently experiences a phase of accelerated expansion
\cite{Riess,perl,abbot}. Understanding the cause of acceleration
of the cosmic expansion has been one of the primarily unsolved
problems in modern cosmology. The component which is responsible
for this acceleration is usually dubbed DE which disclosing its
nature and origin has been on the hottest topics of research in
the past two decades (see \cite{dark1} for a comprehensive
review). The simplest possible candidate for DE is the
cosmological constant, denoted as $\Lambda$ in Einstein's field
equations of general relativity. The resulting model is termed as
$\Lambda$CDM, that is the most acceptable model for describing
both accelerated expansion as well as solving the DM puzzle.
Nevertheless, recent observations indicate a tension with the
$\Lambda$CDM model and reveal that $\Lambda$CDM isn't the best fit
to some data set \cite{luso}. Many alternative theories have been
proposed to modify the matter/energy sector of the Einstein's
field equations. Among these proposals there is a supposition
which plays an important role in finding the nature of DE based on
the holographic principle \cite{susskind1,susskind2,bousso} and is
known as the holographic DE (HDE) \cite{hsu}. This model is based
on the fact that the entropy associated with the boundary is
proportional to the area which was first pointed out by Bekenstein
and Hawking for the black holes \cite{beken1}. According to the
Bekenstein-Hawking area law, the entropy of a black hole is given
by
\begin{equation}
S=\cfrac {A} {4 L_p^{2}},\label{S1}
\end{equation}
where $A$ is the horizon area and $L_p^{2}$ denotes the Planck
area. HDE models have received a lot of attentions in the
literatures \cite{Cohen,Li,huang1,Wang1,Pavon,Wang2,
Setare1,Setare2,Jamil,Shey1,Shey2,Gong,Shey3,Shey4,Shey5,Wang3,Swang,kim}.

Inspired by the Covid-$19$ virus structure, J. D. Barrow argued
that quantum-gravitational effects may deform the geometry of the
black hole horizon leading to an intricate, fractal features
\cite{Barrow}. He discussed that the area law of the black hole
entropy get modified and is given by
\begin{eqnarray}\label{S}
S= \left(\frac{A}{A_{0}}\right)^{1+\delta/2},
\end{eqnarray}
where $A$ is the black hole horizon area and $A_0$ is the Planck
area. The exponent $\delta$ ranges as $0\leq\delta\leq1$ and
represents the amount of the quantum-gravitational deformation
effects. The area law is reproduced in case of $\delta=0$, and
$A_{0}\rightarrow 4L_p^{2}=4G$. On the other hand, the most
intricate and fractal structure of the horizon is obtained by
$\delta=1$. In the cosmological setup, the effects of Barrow
entropy on the cosmic evolution have been investigated from
different viewpoints. For example, modification of the area law
leads to a new HDE model based on Barrow entropy \cite{Emm1,Ana}.
On the other hand, it was recently proven that Barrow entropy as
well as any other known entropy (Tsalis, Renyi, Kaniadakis, etc)
is just sub-case of generalized entropy expression introduced in
\cite{Odin1,Odin2}. Other studies on the cosmological consequences
of the Barrow entropy can be carried out in
\cite{Emm3,Abr1,Mam,Abr2,Bar2,Sri,Das,Sha,Pra,Odin3,Odin4}.

Nowadays, it is a general belief that there is a deep
correspondence between the gravitational field equations and the
laws of thermodynamics (see e.g.
\cite{Jac,Pad,Pad2,Pad3,Fro,Wang0,Ver,Cai5,CaiLM,sheyECFE,SheyLog,SheyPL,
Odin5,Abr} and references therein). In the background of
cosmology, ``gravity-thermodynamics'' conjecture, translates to
the correspondence between Friedmann equations describing the
evolution of the universe and the first law of thermodynamics on
the apparent horizon. It has been confirmed that one can always
rewrite the Friedmann equations in any gravity theory in the form
of the first law of themodynamics on the apparent horizon and vice
versa \cite{CaiKim,Cai2,Sheyk1,Sheyk2}.

In the present work, we are going to investigate HDE when the
entropy associated with the apparent horizon is in the form of
Barrow enetropy given in (\ref{S}). According to the
thermodynamics-gravity conjecture, any modification to the entropy
expression leads to the modified Friedmann equations. The modified
Friedmann equations through Barrow entropy was introduced in
\cite{Emm2,SheBFE1}. A cosmological scenario based on Barrow
entropy was introduce in \cite{SheBFE2}, where it was argued that
the exponent $\delta$ cannot reproduce any term which may play the
role of DE and one still needs to take into account DE component
in the modified Friedmann equations to reproduce the accelerated
universe. On the other hand, the HDE density, which is based on
holographic principle, get modified due to the modification of
entropy. Although, the authors of \cite{Emm1} modify the energy
density of HDE, they still perform their calculations in the setup
of standard Friedmann equations. This is indeed inconsistent with
the thermodynamics-gravity conjecture which states that any
modification to the entropy should modify the field equations of
gravity. Our study differs from \cite{Emm1,Ana}, in that we modify
both energy density of HDE as well as the Friedmann equations
describing the evolution of the universe. Throughout this paper we
set $\kappa_B=1=c=\hbar$, for simplicity.

This paper is outlined as follows. In section II, we study the HDE
in Barrow cosmology when the IR cutoff is the Hubble radius. In
section III, we consider the future event horizon as IR cutoff,
and explore the cosmological consequences of the HDE through
modified cosmology based on Barrow entropy. The last section is
devoted to closing remarks.
%%%%%%%%%%%%%%%%%%%%%%%%%%%%%%%%%%%%%%%%%%%%%%%%%%%%%%%%%%%%%%%%%%%%%%%%%%%%%%%%%%%%%%%%%%%%%%%%%%%%%%%%%%%%%%%%%%%%
\section{HDE with Hubble horizon as IR cutoff}
We consider a homogeneous and isotropic flat universe which
describes by the line element
\begin{equation}
ds^2=-dt^2+a^2(t)\delta_{ij} dx^idx^j,\label{ds}
\end{equation}
where $a(t)$ is the scale factor. Applying Barrow entropy
(\ref{S}) to the holographic framework, one can obtain the HDE
density in the form \cite{Emm1}
\begin{equation}\label{rho}
\rho_{de} =CL^{\delta-2},
\end{equation}
where $L$ is the holographic horizon length and $C$ is a parameter
with dimension $[L]^{-2-\delta}$. For latter convenience, we
choose $C=3c^2M^{2}_{\rm eff}$, where $c^2$ is the holographic
model parameter of order one \cite{Li} and $M^{2}_{\rm eff}$ is
the effective Planck mass which we shall introduce latter. Note
that in the case where $\delta=0$, we have $M^{2}_{\rm
eff}\rightarrow M^2_p$, and the standard HDE density, $\rho_{de}
=3c^2M^{2}/L^{2}$ is restored.

The modified Friedmann equation based on Barrow entropy, in a flat
universe, is given by \cite{SheBFE1,SheBFE2}
\begin{equation} \label{fried0}
H^{2-\delta}=\frac{8 \pi G_{\rm eff}}{3}(\rho_m+\rho_{de}),
\end{equation}
where $H={\dot{a}}/{a}$ is the Hubble parameter, $\rho_m$ and
$\rho_{de}$ are the energy density of pressureless matter and DE,
respectively. Here $G_{\rm eff}$ stands for the effective
Newtonian gravitational constant \cite{SheBFE1}
\begin{equation}\label{Geff}
G_{\rm eff}\equiv \frac{A_0}{4} \left(
\frac{2-\delta}{2+\delta}\right)\left(\frac{A_0}{4\pi
}\right)^{\delta/2}.
\end{equation}
If we define $M^{2}_{\rm eff}= (8 \pi G_{\rm eff})^{-1} $, then
the Friedmann equation (\ref{fried0}) can be written
\begin{equation} \label{fried}
H^{2-\delta}=\frac{1}{3 M^{2}_{\rm eff}}(\rho_m+\rho_{de}),
\end{equation}
Taking the Hubble radius as IR cutoff, $L=H^{-1}$, the HDE density
can be written
\begin{equation}
\rho_{de} =3c^2M^{2}_{\rm eff} H^{2-\delta},\label{rhoH}
\end{equation}
In the remaining part of this section, we shall consider the case
without interaction between dark sectors and the interacting case,
separately.
%%%%%%%%%%%%%%%%%%%%%%%%%%%%%%%%%%%%%%%%%%%%%%%%%%%%%%%%%%%%%%%%%%%%%%%%%%%%%%%%%%%%%%%
\subsection{Non-interacting case}
When two dark components of the universe evolve separately, they
satisfy two independent energy conservation equations,
\begin{eqnarray}\label{cons1}
&&\dot\rho_{de}+3H(1+w_{de})\rho_{de}=0,\\
&&\dot\rho_m+3H\rho_m=0, \label{cons2}
\end{eqnarray}
where $w_{de}=p_{de}/\rho_{de}$ stands for the EoS parameter of
HDE. It is also convenient to introduce the dimensionless density
parameters as
\begin{eqnarray}\label{omegade}
&&\Omega_{de}=\cfrac{\rho_{de}}{\rho_c}=\cfrac{\rho_{de}}{3M^{2}_{\rm eff}H^{2-\delta}},\\
&&\Omega_m =\cfrac{\rho_m}{\rho_c}=\cfrac{\rho_m}{3M^{2}_{\rm
eff}H^{2-\delta}},\label{omegam}
\end{eqnarray}
and thus the Friedmann Eq. (\ref{fried}) can be written
\begin{eqnarray}
&&\Omega_m+\Omega_{de}=1,
\end{eqnarray}
where $\rho_c=3M^{2}_{\rm eff}H^{2-\delta}$ is the effective
critical energy density. Substituting  Eq.(\ref{rhoH}) in
Eq.(\ref{omegade}) one finds $\Omega_{de}=c^2$. Taking the time
derivative of the Friedmann equation (\ref{fried}) and using the
continuity Eqs. (\ref{cons1}) and (\ref{cons2}),  we arrive at
\begin{equation}\label{Hdot}
\frac{\dot{H}}{H^2}=-\frac{3}{2-\delta}(1+\Omega_{de} w_{de}).
\end{equation}
If we take the derivative of Eq. (\ref{rhoH}) respect to cosmic
time, and using Eqs. (\ref{cons1}) and (\ref{Hdot}), we find
$w_{de}=0$. This result is similar to the HDE with Hubble horizon
as IR cutoff in standard cosmology, and cannot lead to an
accelerated universe \cite{Li,Pavon}.
%%%%%%%%%%%%%%%%%%%%%%%%%%%%%%%%%%%%%%%%%%%%%%%%%%%%%%%%%%%%%%%%%%%%%%%%%%%%%%%%
\subsection{Interacting case}
When the two dark sectors of the universe interact with each
other, they do not satisfy the independent conservation equations,
instead they satisfy the semi energy conservation equations as
\begin{eqnarray}\label{consQ1}
&&\dot\rho_{de}+3H(1+w_{de})\rho_{de}=-Q,\\
&&\dot\rho_m+3H\rho_m=Q, \label{consQ2}
\end{eqnarray}
where $Q=3b^2H\rho_{de}$ represents the interaction term. Taking
the time derivative of Eq. (\ref{rhoH}) and using Eqs.
(\ref{Hdot}) and (\ref{consQ1}), we find
\begin{equation}
w_{de}=-\frac{b^2}{1-c^2},\label{wde1}
\end{equation}
which is independent of the exponent $\delta$, and is similar to
the case of standard cosmology \cite{Pavon,Shey3}. In order to
reach $w_{de} < 0$ we should have $c^2<1$. Besides, for
$b^2=1-c^2$, this model mimics the cosmological constant.
Moreover, for $b^2>1-c^2$, this model can cross the phantom line.
The EoS parameter of total energy and matter is defined by
\begin{equation}\label{wtot}
w_{\rm tot}=\Omega_{de} w_{de}=-\frac{b^2 c^2}{1-c^2}.
\end{equation}
It was argued that in modified Barrow cosmology, the condition for
the accelerated expansion reads \cite{SheBFE1}
\begin{equation}\label{wde2}
w_{\rm tot}< -\frac{(1+\delta)}{3}.
\end{equation}
Combining condition (\ref{wde2}) with (\ref{wtot}) implies $3b^2
c^2>(1+\delta)(1-c^2)$ for an accelerated universe. The
deceleration parameter can be easily derived
\begin{eqnarray}
q&=&-1-\frac{\dot{H}}{H^2}=-1+\frac{3}{2-\delta}[1+\Omega_{de}
w_{de}]\nonumber\\
&&=-1+\frac{3}{2-\delta}\left(1-\frac{b^2c^2}{1-c^2}\right).
\end{eqnarray}
which is a constant but depends on $\delta$. In an accelerating
universe $q<0$ which translates to $3b^2 c^2>(1+\delta)(1-c^2)$,
consistent with previous condition (\ref{wde2}).
%%%%%%%%%%%%%%%%%%%%%%%%%%%%%%%%%%%%%%%%%%%%%%%%%%%%%%%%%%%%%%%%%%%%%%%%%%%%%%%%%%%%%%%%%5
\section{HDE with future event horizon as IR cutoff}
In this section we consider the future event horizon as IR cutoff
\cite{Li}
\begin{equation}
{L=R_h=a\int_{t}^{\infty}\cfrac{dt}{a}=a\int_{t}^{\infty}\cfrac{da}{Ha^2}},\label{Lint}
\end{equation}
and therefore the HDE density parameter becomes
\begin{equation}
\Omega_{de}=c^2\left(HR_h\right)^{\delta-2}, \Rightarrow
HR_h=\left(\cfrac{\Omega_{de}}{c^2}\right)^{{1}/{(\delta-2)}}.\label{omega1}
\end{equation}
Taking the time derivative of Eq. (\ref{Lint}) we find
\begin{equation}
\dot{R}_{h}=HR_h-1,\label{Ldot}
\end{equation}
Also, taking the time derivative of energy density $\rho_{de}
=3c^2M^{2}_{\rm eff} R_{h}^{\delta-2}$, yields
\begin{equation}
\dot{\rho}_{de}=3(\delta-2) c^2M^{2}_{\rm eff} \dot{R}_{h}
{R}_{h}^{\delta-3}=\frac{\delta-2}{R_h} \dot{R}_{h}
\rho_{de}.\label{rhodot}
\end{equation}
Again, we shall consider the non-interacting and interacting
cases, separately.
%%%%%%%%%%%%%%%%%%%%%%%%%%%%%%%%%%%%%%%%%%%%%%%%%%%%%%%%%%%%%%%%%%%%%%%%%%%%%%%%%%%
\subsection{Non-interacting case}
\begin{figure}[ht]
\includegraphics[scale=0.8]{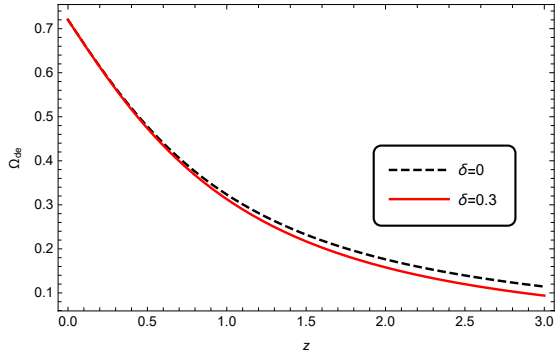}
\caption{Evolution of $\Omega_{de}$ as a function of redshift $z$
for non-interacting HDE. Here we have set $c^2=1$ and
$\Omega_{de,0}=0.72$.}\label{fig1}
\end{figure}

Combining Eqs. (\ref{cons1}), (\ref{Ldot}) and (\ref{rhodot}), one
can obtain
\begin{equation}\label{w}
w_{de}=-\cfrac{1+\delta}{3}+\cfrac{\delta-2}{3}\left(\cfrac{\Omega_{de}}{c^2}\right)^{\frac{1}{2-\delta}},
\end{equation}
for the EoS parameter of HDE. Note that for $\delta=0$, one
restores the EoS parameter of HDE in standard cosmology
\cite{Wang2}. We can also obtain the evolution of density
parameter of HDE. Differentiating $\Omega_{de}$ respect to time
and using relations $\dot\Omega_{de}=H {\Omega^\prime_{de}}$ and
(\ref{Hdot}), we arrive at
\begin{equation}
\Omega^\prime_{de}=\Omega_{de} \left\lbrace
(\delta-2)\left[1-\left(\frac{\Omega_{de}}{c^2}\right)^{\frac{1}{2-\delta}}\right]+
3(1+\Omega_{de}w_{de})\right\rbrace,\label{omegaprime1}
\end{equation}
where the prime denotes derivative with respect to $x=\ln{a}$.
Substituting $w_{de}$ from Eq. (\ref{w}) into the above relation,
yields
\begin{equation}
{\Omega^\prime_{de}}=\Omega_{de}(1-\Omega_{de})\left
\lbrace(1+\delta)+(2-\delta)
\left(\cfrac{\Omega_{de}}{c^2}\right)^{\frac{1}{2\delta}}\right\rbrace.\label{omegaprime2}
\end{equation}
This equation shows the evolution of the density parameter of HDE
in the context of modified Barrow cosmology.  We plot evolution of
$\Omega_{de}$ in Barrow cosmology in Fig. \ref{fig1}. From this
figure we see that at each $z$, the value of $\Omega_{de}$
decreases with increasing $\delta$. Besides, the difference
between $\Omega_{de}$ in standard cosmology and modified Barrow
cosmology increases at higher redshifts.
\begin{figure}[ht]
\centering
\includegraphics[scale=0.8]{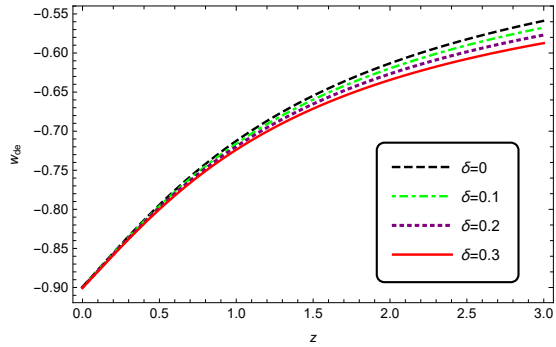}
\caption{Evolution of $w_{de}$ as a function of redshift $z$ for
non-interacting HDE. Here we set $c^2=1$, $M^{2}_{\rm
eff}=1$.}\label{fig2}
\end{figure}
The evolution of EoS parameter $w_{de}$ for HDE versus $z$ is
plotted in Fig. \ref{fig2} for different values of $\delta$. We
see that at the present time, the values of $w_{de}$ is
independent of $\delta$, while its behavior differs for early
times. For $z=0$ we have $w_{de}\approx-0.9$, and increasing
$\delta$ will decrease $w_{de}$ in all times. For example, for
$\delta=0.3$ at $z=0$ we have $w_{de}=-0.91$ which is closer than
the standard model to observational data. Therefore, in modified
Barrow cosmology, $w_{de}$ of HDE at the present time, is
compatible with observations.

\begin{figure}[ht]
\centering
\includegraphics[scale=0.8]{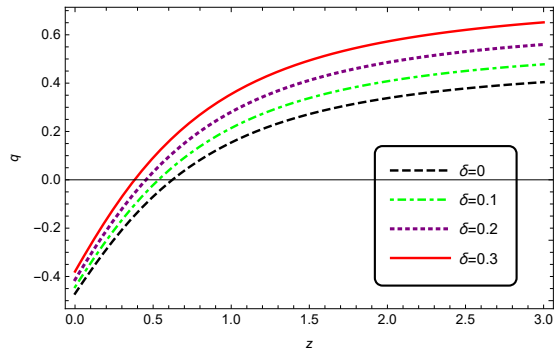}
\caption{The behaviour of the deceleration parameter $q$ as a
function of redshift $z$ for non-interacting HDE in Barrow
cosmology, where we set $c^2=1$.}\label{fig3}
\end{figure}

Next, we examine the deceleration parameter, $q=-\ddot{a}/(aH^2)$.
Using Eqs. (\ref{Hdot}) and (\ref{w}), one finds
\begin{eqnarray}
q&=&-1-\frac{\dot
H}{H^2}=-1+\cfrac{3}{2-\delta}\left(1+\Omega_{de}w_{de}\right)\nonumber\\
&=&\cfrac{1+\delta}{2-\delta}(1-\Omega_{de})-\Omega_{de}
\left(\cfrac{\Omega_{de}}{c^2}\right)^\frac{1}{2-\delta}.\label{qflat}
\end{eqnarray}
The evolutionary behavior of the deceleration parameter is plotted
in Fig. \ref{fig3}. From this figure we see that with increasing
the Barrow exponent $\delta$, the transition from deceleration to
acceleration takes place in lower redshift.
%%%%%%%%%%%%%%%%%%%%%%%%%%%%%%%%%%%%%%%%%%%%%%%%%%%%%%%%%%%%%%%%%%%%%%%%%%%%%%%%%%%%%%%%%%%%%%%%%%%%%%%
\subsection{Interacting case}
\begin{figure}[ht]
\centering
\includegraphics[scale=0.8]{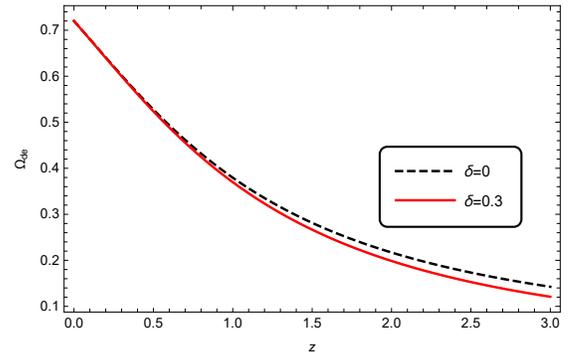}
\caption{Evolution of $\Omega_{de}$ as a function of redshift $z$
for interacting HDE in Barrow cosmology. Here we take $c^2=1$,
$M^{2}_{\rm eff}=1$ and $b^2=0.1$.}\label{fig5}
\end{figure}

Next, we investigate the case for FRW universe filled with HDE and
DM exchanging energy. We consider the future event horizon as IR
cutoff. In interacting case the conservation equations are in the
form of Eqs.(\ref{consQ1}) and (\ref{consQ2}). Following the
method of the previous section, we can find the EoS parameter as
\begin{equation}
w_{de}=-b^2-\cfrac{\delta+1}{3}+\cfrac{\delta-2}{3}
\left(\cfrac{c^2}{\Omega_{de}}\right)^\frac{1}{\delta-2}.\label{Wint}
\end{equation}
In this case the evolutionary equation for $\Omega_{de}$ becomes
\begin{eqnarray}
{\Omega^\prime_{de}}&=&\Omega_{de}(1-\Omega_{de})\left
\lbrace(1+\delta)+(2-\delta)
\left(\cfrac{\Omega_{de}}{c^2}\right)^{\frac{1}{2\delta}}\right\rbrace\nonumber\\
&&-3b^2 \Omega_{de}^2,\label{omegaprimeint}
\end{eqnarray}
while the deceleration parameter reads
\begin{eqnarray}
&&q=-1-\frac{\dot
H}{H^2}=-1+\cfrac{3}{2-\delta}\left(1+\Omega_{de}w_{de}\right)\nonumber\\
&=&\cfrac{1+\delta}{2-\delta}(1-\Omega_{de})-\Omega_{de}
\left(\cfrac{\Omega_{de}}{c^2}\right)^\frac{1}{2-\delta}-3\frac{b^2\Omega_{de}}{2-\delta}.\label{qflat}
\end{eqnarray}
We plot the evolution of $\Omega_{de}$ for different values of
$\delta$ in Fig. \ref{fig5}.

\begin{figure}[ht]
\centering
\includegraphics[scale=0.8]{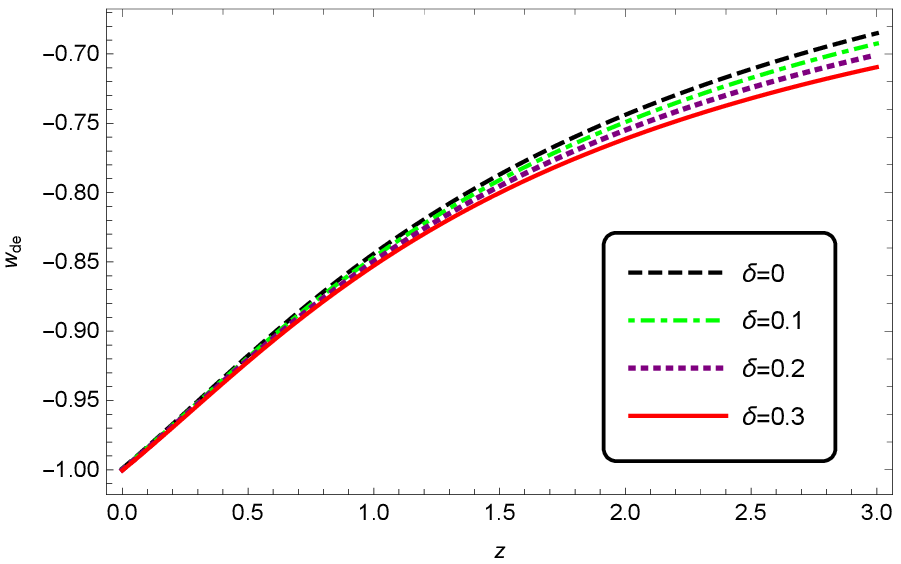}
\caption{The evolution of EoS parameter $w_{de}$ as a function of
redshift $z$ for interacting HDE. Here we take $c^2=1$,
$M^{2}_{\rm eff}=1$, $b^2=0.1$.}\label{fig6}
\end{figure}

\begin{figure}[ht]
\centering
\includegraphics[scale=0.8]{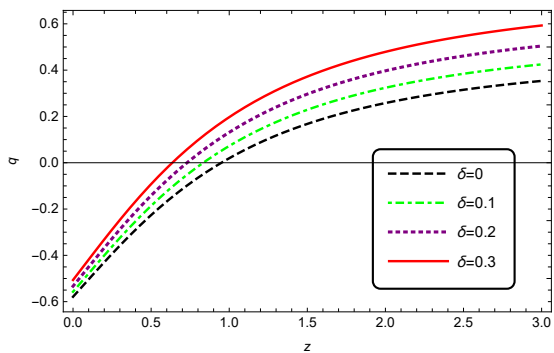}
\caption{Evolution of the deceleration parameter $q$ as a function
of redshift $z$ for interacting HDE. We have taken $c^2=1$,
$M^{2}_{\rm eff}=1$, $b^2=0.1$.}\label{fig7}
\end{figure}

\begin{figure}[ht]
\centering
\includegraphics[scale=0.8]{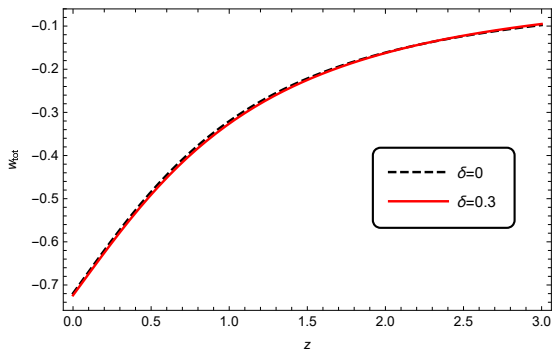}
\caption{Evolution of $w_{tot}$ as a function of redshift $z$ for
interacting HDE, where $c^2=1$, $M^{2}_{\rm eff}=1$,
$b^2=0.1$.}\label{fig8}
\end{figure}

\begin{table}[t]
\begin{center}
\begin{tabular}{c|c|c|c|}
\cline{2-4}
 $w_{de}$ & $b^2=0.03$ &$b^2=0.06$ & $b^2=0.1$ \\
\hline
\multicolumn{1}{|c|}{$\delta=0$} & $-0.92909$ & $-0.95909$ & $-0.99909$ \\
\hline
\multicolumn{1}{|c|}{$\delta=0.1$} & $-0.92944$ & $-0.95944$ & $-0.99944$\\
\hline
\multicolumn{1}{|c|}{$\delta=0.2$} & $-0.92991$ &$-0.95991$ & $-0.99991$ \\
\hline
\multicolumn{1}{|c|}{$\delta=0.3$}  & $-0.93049$ & $-0.96049$ & $-1.00049$\\
\hline
\end{tabular}
\caption{Numerical results for EoS parameter of interacting HDE at
the present time, $z=0$.}\label{table1}
\end{center}
\end{table}
We can also plot the EoS and the deceleration parameter for the
interacting HDE in modified Barrow cosmology. From Fig. \ref{fig6}
we see that with increasing $\delta$, the EoS parameter decreases
as well. We also observe from Fig. \ref{fig7} that with increasing
the Barrow exponent $\delta$ the transition from deceleration to
acceleration occurs at lower redshift.
Evolution of the total EoS parameter for different values of
$\delta$ in Fig. \ref{fig8} indicates that there is no significant
difference between $w_{tot}$ of HDE in standard cosmology
($\delta=0$) and in modified Barrow cosmology ($\delta\neq0$).

\begin{figure}[ht]
\centering
\includegraphics[scale=0.8]{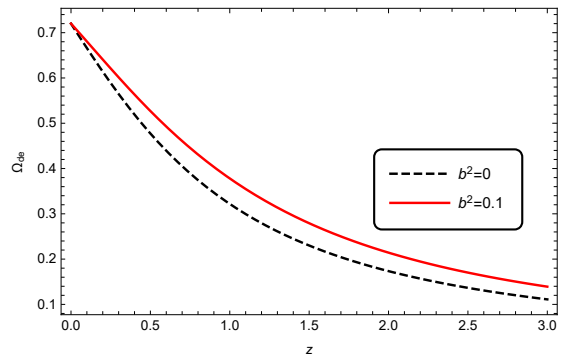}
\caption{Evolution of  $\Omega_{de}$ versus $z$ for modified
Barrow HDE, where $c^2=1$, $M^{2}_{\rm eff}=1$,
$\delta=0.2$.}\label{fig9}
\end{figure}

By fixing the Barrow component $\delta$ and changing the values of
$b^2$ we can reveal the effects of interaction term on the
cosmological behavior of our model. In Fig. \ref{fig9}, we plot
the evolution of $\Omega_{de}$ for interacting HDE in modified
Barrow cosmology. It is seen that at each time, the value of
$\Omega_{de}$ is higher compared to standard cosmology.

\begin{figure}[ht]
\centering
\includegraphics[scale=0.8]{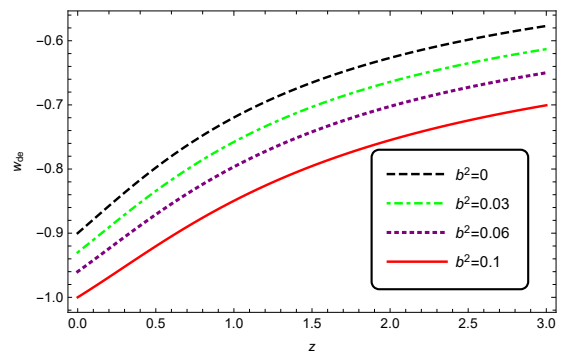}
\caption{Evolution of EoS parameter $w_{de}$ as a function of
redshift $z$ for interacting HDE. Here we take $c^2=1$,
$M^{2}_{\rm eff}=1$, and $\delta=0.2$.}\label{fig10}
\end{figure}

\begin{figure}[ht]
\includegraphics[scale=0.8]{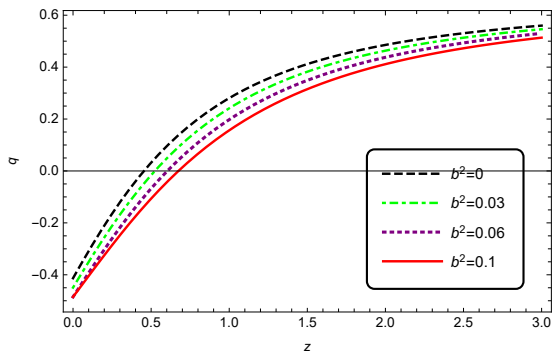}
\caption{Evolution of the deceleration parameter $q$ as a function
of redshift $z$ for interacting HDE in modified Barrow cosmology.
Here we have taken $c^2=1$, $M^{2}_{\rm eff}=1$, and
$\delta=0.2$.}\label{fig11}
\end{figure}

\begin{figure}[ht]
\includegraphics[scale=0.8]{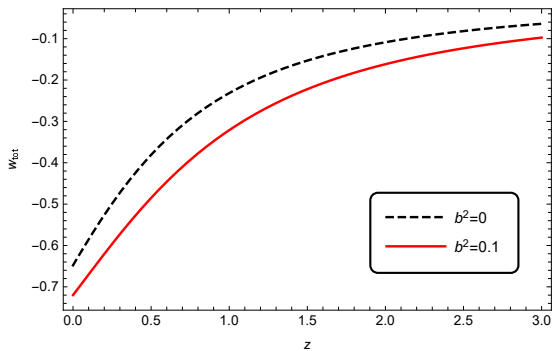}
\caption{Evolution of $w_{tot}$ as a function of redshift $z$ for
interacting HDE in modified Barrow cosmology. Here we have taken
$c^2=1$, $M^{2}_{\rm eff}=1$, and $\delta=0.2$.}\label{fig12}
\end{figure}

In Fig. \ref{fig10} we can see that $w_{de}$ of the interacting
HDE decreases with increasing coupling constant $b^2$. For
$\delta=0.2$ and $b^2=0.1$ we see that $w_{de}$ crosses the
phantom regime.
%%%%%%%%%%%%%%%%%%%%%%%%%%%%%%%%%%%%%%%%%%%%%%%%%%%%%%%%%%%%%%%%%%%%
As one can see in Fig. \ref{fig11}, changing $b^2$ has affected on
the phase transition of interacting HDE in Barrow cosmology. We
observe that unlike the effects of adding the values of $\delta$
on deceleration parameter, increasing $b^2$ will shift the
transition to higher redshifts and it means that the expansion of
the universe changed from decelerating phase to an accelerating
phase in much earlier times. By comparing the effects of changing
$\delta$ and $b^2$, given by Figs. \ref{fig7} and \ref{fig11}, we
can see that the best values for the parameters are in the range
$0.06<b^2<0.1$ and $0.2<\delta<0.3$.
Again by using our previous results for $\Omega_{de}$ and
$w_{de}$, we can plot the total EoS parameter, for different
values of $b^2$ in case of interacting HDE which is shown in Fig.
\ref{fig12}. Adding coupling constant will decrease $w_{tot}$
significantly during all times and reveals the difference between
interacting and non-interacting HDE in the framework of Barrow
cosmology.
%%%%%%%%%%%%%%%%%%%%%%%%%%%%%%%%%%%%%%%%%%%%%%%%%%%%%%%%%%%%%%%%%%%%%%%%%%%%%%%%%%%%%%%%%%%%%%%%%%%%%%%%%%%%%%%
\section{Closing remarks}
According to thermodynamics-gravity conjecture, any modification
to the entropy modifies the energy density of HDE as well as the
Friedmann equations. Based on this and using the modified
Friedmann equations, we investigated the HDE when the entropy is
in the form of Barrow entropy. We first have taken the Hubble
radius as IR cutoff and showed that the accelerated universe can
be achieved only in the interacting case. This behavior is similar
to HDE in standard cosmology with Hubble cutoff. We then
considered the future event horizon as IR cutoff and investigated
both non-interacting and interacting HDE in a flat universe.

We examined the effects of Barrow exponent $\delta$ on the
cosmological evolution of HDE. We observed that for $\delta=0$ the
EoS parameter of non-interacting HDE lies completely in the
quintessence regime, while for the interacting case it can cross
the phantom line. In both cases, with increasing $\delta$, the EoS
parameter $w_{de}$ decreases at any time. Another interesting
result we found in this work is that the presence of $\delta$ can
change the time of phase transition of the universe from
deceleration to the acceleration. Indeed, with increasing $\delta$
the phase transition occurs at lower redshifts. This behavior is
seen for both non-interacting and interacting cases. On the other
hand, for a fixed value of $\delta$, if we increase the coupling
constant of interaction term, the transition occurs in higher
redshift.

To sum up, the incorporation of modified Friedmann equations to
HDE improves the phenomenology comparing to the standard Friedmann
equation while keeping the Barrow exponent $\delta$ to smaller
values. This is an advantage of this scenario, since in a more
realistic case we expect the Barrow exponent to be closer to the
standard Bekenstein-Hawking value and have the results compatible
with observations.
%%%%%%%%%%%%%%%%%%%%%%%%%%%%%%%%%%%%%%%%%%%%%%%%%%%%%%%%%%%%%%%%%%%%
\acknowledgments{We thank Mahya Mohammadi for helpful discussion
and valuable comments.}
%%%%%%%%%%%%%%%%%%%%%%%%%%%%%%%%%%%%%%%%%%%%%%%%%%%%%%%%%%%%%%%%%%%%%%%%%%%%%%%%%%%%%%%%%%%%%%%%%%%%%%%%

\end{document}